# Instant Computing
## A new computation paradigm
### by Hans-Rudolf Thomann
### hr@thomannconsulting.ch

### 2008-01-12

| | |
|---|---|
| **Author** | Hans-Rudolf Thomann |
| **Version** | 1.0.1 |
| **Date** | 12 January 2008 |
| **Status** | Final version |
| **Action** | Read the preface and at least the changed sections |
| **Distribution** | arXiv.org cs/0610114 |

# CONTENTS

## Abstract

Voltage peaks on a conventional computer's power lines allow for the well-known dangerous DPA attacks. We show that measurement of a quantum computer's transient state during a computational step reveals information about a complete computation of arbitrary length, which can be extracted by repeated probing, if the computer is suitably programmed.

That's instant computing, a mode of operation of quantum computers, which is the subject of this paper. Hamiltonian automata (HA), modeling physical implementations of Quantum Turing Machines (QTM), are defined in the framework of non-relativistic Quantum Mechanics (QM), and the instant computing mode of operation is specified and analyzed. When random sampling HA from the set of all implementations conforming to this specification measurement of the transient state after one half machine cycle reveals the computation result with probability $O(1)$. With an average constant number of trials any computation result is obtained.

For any total or partial recursive function, instant computing recognizes arguments of length $n$ lying in the domain of definition and yields their function value with arbitrarily small error probability in probabilistic linear time $O_p(n)$. This implies recognition of (not necessarily recursively enumerable) complements of recursively enumerable sets and a solution of the halting problem.

As a byproduct, physical processes of arbitrary Turing-computational time complexity as well as Turing-uncomputable processes are obtained. These results refute popular generalizations of the Church-Turing Thesis (CT), establish a new computing paradigm, provide a new approach to the design of computing devices and physical tools, and may lead to a new understanding of certain natural phenomena.

Keywords: Cryptology, Quantum Computing, Computational Complexity, Theory of Computation, Halting Problem, Church-Turing Thesis

## Preface to Version 1.0

The final draft (version 0.9) has been posted for review since October 2006. I thank the reviewers for their qualified comments. The review confirmed Instant Computing as described in chapter 2, while challenging the equal distribution assumption in section 3.2.1. These concerns are fully addressed and (hopefully) resolved by strengthening and generalizing the statistical analysis. All comments received are considered in this consolidated final version.

The most significant change has been made in section 3.2.1 where the statistics of implementations are analyzed. The former assumption of an underlying equal distribution has been replaced by Lemma 3.1 stating that the $y_k$ are i.i.d. with an arbitrary underlying distribution. Theorem 3 still holds in this much more general setting: Random implementations are likely to exhibit instant computing. I hope this generalization and the arguments added are sufficient evidence for the likelihood of instant computing realizability.

Some reviewers interpreted our statements about the existence of physical implementations from a physicists or engineer's point of view, whereas they should be read as claims of the mathematical existence of some Hamiltonian operators. Furthermore, it did not become sufficiently clear that our results on realizability are conditional, under the proviso that QTMs indeed are realizable! This has been clarified in the introduction and throughout the paper.

Finally, we were asked to explain how the future computation results could be anticipated at one half machine cycle. The answer is, that anticipation arises from the fact that quantum states are superpositions of eigenstates, and these are superpositions of all states occurring in the course of a computation. So anticipation is inherent in quantum evolution.

Besides that, various typos and inaccuracies but no major flaws or bugs have been reported and are corrected in this version. The abstract and introduction have been adjusted to the new theorem 3.

The main deficiency of this publication however remains: It is overloaded. Too many aspects are squeezed into 30 plus pages, and still not treated extensively but only sketched. This makes the judgment and understanding difficult, particularly for readers whose background are quantum computing mainstream concepts and the related mathematics. Therefore a series of papers will be produced, each one devoted to a single aspect. You are kindly invited to watch out for them and to enjoy their reading.

Zürich, 7 January 2008

Hans-Rudolf Thomann

# 0 Introduction

## 0.1 Question

Today's computers essentially are complex CMOS transistor circuits driven by the clock through their computational states. During state transition, some number of transistors is switching. Whereas in a stable 0 or 1 state CMOS transistors have a high impedance, while switching they almost shortcut. This causes (negative) voltage peaks on the power lines proportional to the number of transistors switching. [Ko99] demonstrated how to use such power measurements to reveal information about the computer's internal state and to mount a most dangerous class of side-channel attacks on computers and cryptographic algorithms, named Differential Power Analysis (DPA).

Tomorrow's computers may well be quantum computers maintaining a coherent state, driven from state to state by some Hamiltonian. They will as well perform operations step by step, proceeding from one computational state to the next one. What do you see when measuring their transient state? Though quantum computers are not yet on stocks at Radio Shack, this question can be analyzed. This reveals an effect with astounding applications to computing.

## 0.2 Basic Idea

Due to coherence of quantum states, the transient state can be expected to be a superposition of a certain number of states occurring in the course of a computation. What states do occur depends on the program. Now program the computer such that it counts its own steps and, as it reaches the result, leaves it on the working tape for a multiple of the step count, and thereafter reverses the computation (as described in [Be73]). The analysis proves: With probability $O(1)$ the result appears in the transient state after one half machine cycle.

All we have to do is to measure the program state after one half machine cycle. If the computation reaches a result, then with probability $O(1)$ we obtain the result. If not, then we prepare the initial state again, restart the quantum computer and repeat the measurement. After average $O(1)$ trials we are finished.

That's instant computing[1], a particular mode of operation of quantum computers, and neither a new type of quantum computers nor a new definition of computation. But it yields the result of arbitrarily complex computations after one half machine cycle. Therefore it is more powerful than any classical or quantum computer in common mode of operation.

Even some Turing-uncomputable problems are solvable by instant computing. This is, because instant computing is a physical computation process, exploiting quantum interference, operating in continuous space-time, whereas Turing machines operate in an abstract, discrete state space and step time.

Impossible? Prohibited by the Church-Turing thesis? This popular misbelief is clarified in the final chapter. Read the next chapters and convince yourself that instant computing has solid fundament in the laws of Quantum Mechanics. If some day QTMs can be realized, then instant computing is likely to become reality.

---

[1] An allusion both to instantaneous response as well as to instant coffee!

### 0.3    Outline of Results

Instant computation is a probabilistic computation procedure operating on Hamiltonian Automata (HA), i.e. abstract quantum-physical systems implementing Quantum Turing Machines (QTM) emulating Deterministic Turing Machines (DTMs).

From a HA implementation of a DTM M computing some Turing-computable function $f$ the function evaluation result is obtained, in half the machine cycle time, $\frac{T}{2}$, with probability $p$. This construction is applicable to any Turing-computable $f$, partial or total (proposition 2.3). The probability $p$ and time $T$ do not depend on the problem size.

Theorem 1 states that, on arguments in the DTM's domain of definition, the function value is obtained in probabilistic linear time $O_p(n)$ (see [Pr59] for the Chernov-Pratt theory of stochastic order). Arguments, on which the DTM does not halt, are recognized with bounded error in probabilistic linear time $O_p(n)$. Problems in $NEXPPOLY$ are solvable with certainty in probabilistic polynomial time $O_p(poly)$.

Any recursively enumerable set and its (not necessarily recursively enumerable) complement thus become decidable. The existence of physical processes of arbitrary Turing-computational time complexity, as well as of Turing-uncomputable physical processes, is established (proposition 4.1).

Instant computation only assumes the capability to physically implement Quantum Turing Machines (QTMs) and to operate them: Prepare an initial state, start the machine and measure the state after time $\frac{T}{2}$, where $T$ is the machine cycle time! Finite resources, operations and measurement precision are sufficient. No account to real numbers and real computation is made, no particularly high precision measurements or any other capabilities beyond QTM technology are required.

The (mathematical) existence of infinitely many Hamiltonian operators exhibiting the behavior required for instant computation is established (proposition 2.4). HA implementing universal Turing Machines and taking only finite physical resources do exist, in the mathematical sense (theorem 2). When randomly selecting a HA from set of all efficient (see definition 1.4.b) implementations, with probability $1 - o_p(1)$, the computation result has probability amplitude $O(1)$ in the transient state at one half machine cycle (theorem 3). If some day QTMs can be realized, then theorem 3 shows that they are likely to exhibit instant computing.

The demonstrated power of physical processing relies on the fact that quantum-mechanical systems are, in contrast to Turing Machines (TM), not finitely defined in the sense of [Ga80]. Unlike TMs acting on discrete configurations in discrete step time governed by a finite program, QTMs operate in continuous space-time driven by a Hamiltonian H. H has an infinite effect on the physical state, and the series expansion $e^{-iHt} = \sum \frac{(-iHt)^n}{n!}$ comprises, for any t, arbitrary powers of H, i.e. arbitrary compositions of its basic effect. The state of a physical system evolves thus for any $t > 0$ in a superposition of infinitely many base states. The eigenstates are superpositions of all states occurring in the course of a computation. They anticipate the computation result if the machine is suitably programmed.   From this instant computation takes profit.

Our result is obtained in abstract terms of Mathematical Physics based on the axioms of non-relativistic Quantum Mechanics. Though experimentalists are skilled to approximately construct even bizarre Hamiltonians, it is worth investigating possible obstacles. One is the Hamiltonian operator type, which we take to be any self-adjoint operator, but is constrained by major physical models to Schrödinger form $H = -\Delta + V$. Our mathematical method to construct implementations of DTMs is not applicable to this sub-class. Their existence (in the mathematical sense) as well as implementations in the framework of Quantum Field Theory (QFT) are open problems.

## 0.4    Organisation of the Paper

Section 1 presents definitions and basic properties of as well as some important facts about Quantum Turing Machines (QTM), quantum-physical systems, HA, implementations and physical time complexity. Section 2 defines instant computation and specific implementations of DTMs, proves the (mathematical) existence of suitable implementations and establishes the main theorem (theorem 1). Section 3 investigates implementation complexity (theorem 2) and the statistical properties of the set of implementations (theorem 3), and points out some properties of Schrödinger Operators making implementations more challenging. The final section summarizes the results and takes some philosophical conclusions.

As this paper is about TMs, particularly QTMs, their implementations into quantum-physical systems and the resulting computational power, we assume some familiarity with automata theory, computational complexity, Quantum computation and mathematical physics of Quantum Mechanics, and their terminology.

# 1 Definitions and Basic Properties

We recall here the definition and some important properties of QTMs and physical systems, define HA and QTM implementations and investigate their properties. Terminology is indicated by italic type.

## 1.1 QTMs

Recalling from [BV93], a QTM M is given by a unitary *evolution operator* $U_M$, a *configuration space* C, a Hilbert space, and a *configuration* c. The base vectors of C derive from conventional tape inscriptions as follows: To each element of the alphabet, a base vector is associated. The base configuration is the tensor product of the vectors corresponding to the contents of each field of the working tape.

$U_M$ is generated by the *finite state control*, and thus *finitely defined* in the sense of [Ga80]. $U_M$ is *elementary*, if application of $U_M$ effects one single computational step of the QTM M, i.e. one elementary machine operation. Though there are prominent examples such as [DJ92] and [Gr96], where the QTM is specified in non-elementary form, any QTM possesses such an elementary evolution operator.

Deterministic TMs (DTM), brought into reversible form (see [Be73, 82, 93, 95] for reversible computation), can be implemented in QTMs. Their evolution operators are permutation operators, those of general QTMs more general unitary operators. Subsequently, we will not strongly distinguish DTMs from their QTM implementations.

General QTMs may be conceived as having the capability to perform deterministic computations in quantum-parallel mode of operation, branching into quantum-parallel computational threads which eventually interfere.

The *pure input configurations* of a QTM solving a problem with size parameter $n$ are kets of form $|x_1, ..., x_n, n\rangle$. Thus inputs of different size or value are orthogonal.

Given a pure input $c_0$, an elementary $U_M$ generates the *computational sequence* $c_n = U_M^n c_0$. The inner product $(c_j, c_k) = \alpha_{j-k}$ depends only on the difference of j and k. If M is a DTM, then $\alpha_{j-k} = \delta_{(j-k) \bmod p}$, where $p \leq \infty$ is the *period* of the computation. I.e. the $c_k$ are mutually orthogonal. The computational sequence spans the *computation space* $S_{C,c} = Span(\{c_n | n \in \mathbb{Z}\})$ of dimension $p$.

The mutual orthogonality of the computational states of DTMs eases the analysis. Therefore we focus on DTMs in our specification of instant computing.

QTMs cannot halt or make output. Instead they reach certain states labeled *result states*. The *computational time complexity* measure of QTMs is the number of steps from start to result. If necessary, we will denote this measure by Turing time complexity, to distinguish it from HA model time complexity. Note that, as different branches generally have different time complexity, even when starting from a pure input configuration, the step number will rather be a probability distribution than a function. Notice that we take as *problem size*, $n$, the sum of input and output length.

Definition 1.1
  a) The *domain of definition* of a QTM M is the set of those pure inputs on which M eventually reaches a result state.
  b) A QTM M is *periodic on input* $c$ if, when starting from $c$, it returns in a finite number of steps back to $c$ and thus loops ad infinitum.

## 1.2 Physical Systems

Recalling from [LL77, SR80], a *quantum-physical system* S is given by a self-adjoint operator H, the *Hamiltonian*, operating on a Hilbert space Q, and a state q on the surface of the unit ball. H generates the unitary *evolution operator*

$$U_t = e^{\frac{-iHt}{\hbar}},$$

generating the *orbit*

$$q_t = U_t q_0,$$

spanning the *orbital space*

$$S_{S,q} = Span(\{q_t | t \in \mathbb{R}\}).$$

All *moments* of H

$$(q, H^n q) \overset{def}{=} \langle H^n \rangle \ (n = 0, 1, \dots)$$

are constants of motion, i.e. constant on any orbit. Particularly, $\|q\|_2$ and $E = \langle H \rangle$ are time invariant. $q$ is a *physical state*, iff $\langle H \rangle < \infty$.

H has a *spectral representation*

$$(q, H^n q) = \int \lambda \, d\mu_q(\lambda).$$

As quantum systems obey *time reversal* (see [LL77, §§18 and 60]), H and thus $U_t$ has a system of real-valued eigenfunctions. Furthermore, the multiplicity of eigenvalues is constrained (see [LL77, RS80]).

The time function
(1.1)                          $d\mu_q^\wedge(t) = (q_0, q_t)$

is the Fourier transform of the *spectral measure* $d\mu_q$. It satisfies $\left| d\mu_q^\wedge \, t \right| \le 1$ for all t and $d\mu_q^\wedge(0) = 1$. If for some $t \ne 0$: $\left| d\mu_q^\wedge(t) = 1 \right|$, then it is (up to phase) periodic with period t.

By the semi-group property of $U_t$,

$$(q_s, q_{s+t}) = d\mu_q^\wedge(t),$$

independent of $s$.

## 1.3 Hamiltonian Automata

A HA P is given by a physical system S and a complete set O of *observables*, each of them with finite point spectrum. Computation by a HA is accomplished by first preparing a state in terms of O, then letting act the Hamiltonian H for a time span t and then to measure the state in terms of O.

All our results hold for arbitrary choice of O.

### 1.4    Physical Implementations

### 1.4.1    Definition

A *physical implementation* of a QTM M with elementary evolution operator $U_M$ is given by a HA P and a subset O' of its observable set, related to M as follows:

- Each element of O' has $|A|$ eigenvectors, where A is M's alphabet.
- A unitary mapping $\varphi$ between C and a (proper or improper) subspace Q' of Q is induced by O', by associating to any base configuration c that uniquely determined physical state in Q', where observable $o_k$ is prepared to the same value as working tape field k. In turn, any state of Q has a uniquely determined image in C obtained by successively measuring the observables.
- For some $T > 0$ (the *machine cycle time*), $\varphi * U_T \varphi = U_M$.
- Preparation or measurement of observable $o_k$ after observable $o_j$ takes time $O(k - j)$.

An immediate consequence of this definition is that

(1.2)
$$\alpha_{j-k} = (c_j, c_k) = d\mu_q^\wedge (T(j - k)).$$

Computation by a HA is accomplished by first preparing a state in terms of O, then letting act the Hamiltonian H for a time span t, and then to measure the state in terms of O. See [Bu97] for theory of preparation and measurement.

The main difference between QTMs and QTM implementations is that the latter operate in *continuous time*, the former in discrete *step time*. I.e. the state of a QTM is only defined at integer values of t, that of a HA also for intermediate values of t. As we will see later on, evolution in continuous time exhibits effects inaccessible to QTMs.

Our definition of implementations makes use only of observables with finite range, in terms of which inputs are easily prepared and outputs measured. All our results are independent of the choice of O, or equivalently, $\varphi$. This choice relates $U_M$ and $U_T$ and, for any DTM M, constrains the Hamiltonian, H, particularly its projection-valued measure, without defining it completely, thus leaving freedom for uncountably many implementations.

### 1.4.2    Spectrum

For some self-adjoint operator G,
$$U_T = e^{-iG}.$$

The spectral decomposition of G
$$(q, Gq) = \int_{\kappa=0}^{2\pi} \kappa \, d\nu_q(\kappa),$$

is related to H's spectral decomposition by
$$d\nu_q(\kappa) = \sum_{\kappa = \frac{\lambda T}{\hbar} \bmod 2\pi} d\mu_q(\lambda),$$

which we express by writing

(1.3)
$$G = \frac{HT}{\hbar} \bmod 2\pi.$$

Proposition 1.2

Let P be an implementation of a DTM M.

a) A computation $c_k$ has period $p$ iff the corresponding orbit $q_t$ has period $pT$. The spectrum of G and of H (restricted to this orbit) is pure point iff $p < \infty$, absolutely continuous otherwise.

b) If orbit $q_t$ has period $p < \infty$, then the eigenvalues of G (restricted to this orbit) are

$\kappa_j = \dfrac{2\pi j}{p}$ $(j = 0, \dots, p-1)$ and $\mu(\{\kappa_j\}) = \dfrac{1}{p}$. The eigenvalues of H satisfy

$\dfrac{\lambda T}{\hbar} = 0 \bmod \dfrac{2\pi}{p}$.

(1.4)
$$f_j = \frac{1}{\sqrt{p}} \sum_{k=0}^{p-1} e^{2\pi i \frac{jk}{p}} q_k$$

are eigenvectors. If the orbital space S has dimension $p$, then they are a basis of S. If the dimension of S is $> p$, then there is a basis which they are part of.

c) If $p = \infty$, then $G$ (restricted to this orbit) has spectral measure $d\mu(\kappa) = \dfrac{d\kappa}{2\pi}[0 \leq \kappa \leq 2\pi]$.

d) The states $q_{T(k \bmod p)}$ $(k \in \mathbb{Z},\ 0 < p \leq \infty)$ are part of a basis of the orbital space S. If S has dimension $p < \infty$, then they are a basis of S.

Proof:

Recall that, as M is a DTM, the $q_{T(k \bmod p)}$ $(k \in \mathbb{Z},\ 0 < p \leq \infty)$ are mutually orthogonal.

a) The computation and the orbit are related by the unitary mapping $\varphi$. $U_M$ is a bounded operator with spectrum on the complex unit circle. The spectral measure is uniquely determined by $(f(U_M)q_0, q_0)$ $(f \epsilon C^\infty(\mathbb{C}))$. These functions in turn are uniquely determined by the monomials $z^k$ which due to mutual orthogonality satisfy $(U_m^k q_0, q_0) = (q_{kT}, q_0) = \delta_{k,0}$. Thus the spectral measure of $U_M$ is equally distributed, in the periodic case over the unit roots, in the infinite case over the whole complex unit circle. Mixed spectrum cannot occur, as in the periodic case the initial state must recur, whereas in the aperiodic case $d\mu_q^\wedge(t) \epsilon o(1)$.

b) The eigenvalues and their spectral measure are obtained from formula 1.3 and part a). Formula 1.4 follows again from the mutual orthogonality of the $q_k$.

c) This follows from part a).

d) From mutual orthogonality.                                            q.e.d.

### 1.4.3 Existence

To formulate the (mathematical) existence results we first define computability relative to an arbitrarily chosen basis B. This model avoids reference to and the intricacies of real-computation. Proposition 1.3 below relates B, $\varphi$ and O. All our results are independent of their choice.

Definition 1.2

Let $B = \{b_k\}$ be an orthogonal basis of a Hilbert space Q.

a) A state $q \in Q$ is *computable from B*, if there is a QTM M which for any computable $\varepsilon > 0$ computes a sequence $\alpha_k$ of length $m_\varepsilon$, such that $\left\| q - \sum_{k=0}^{m_\varepsilon} \alpha_k b_k \right\|_2 < \varepsilon$.

b) An orthogonal operator U is *computable from B*, if for any n, the restriction of U to problems of size n is computable from B, i.e. there are QTMs M and N, such that M, given a state $q$ computable from B by N, problem size $n \in \mathbb{N}$, computable $t \in \mathbb{R}, \varepsilon > 0$ yields a state $q_{t,\varepsilon}$ computable from B by N.

c) A self-adjoint operator H is *enumerable from B*, if there is a recursively enumerable sequence $\lambda_k \epsilon \mathbb{R} \, (k = 0,1,\dots)$ such that $\lambda_k$ is the eigenvalue of H for eigenvector $b_k \in B$.

## Proposition 1.3

a) a) Let M be any DTM, Q any Hilbert space and $\varphi$ any unitary injection from C into Q. Then M has uncountably many physical implementations with state space Q and basis B.

b) b) Part a) remains valid if $\varphi$ is required to map pure inputs of M on physical states, i.e. states with finite energy.

Proof:

a) Let $\varphi$ be any unitary injection from C into Q. We first construct a set O of observables, by reduction of measurement in space Q to measurement in space C:

$$o_k = \varphi \omega_k \varphi *,$$

where $\omega_k$ denotes measurement in space C of field number k of M's working tape. Each operator has spectrum $\lambda_k = k \quad k = 0,\dots,|A| - 1$ , where A is M's alphabet. We obtain in the same way

$$U_T = \varphi U_M \varphi *$$

Now define the basis B by $b_k = \varphi c_k$, where $c_k$ denotes a base state in configuration space C. On periodic orbits, G has pure point spectrum, and eigenvectors and eigenvalues of G and $U_T$ are computable from B by formula 1.4. On a-periodic orbits, G has continuous spectrum, and approximate eigenfunctions are obtained from formula 1.4 for large n.

G is related by formula 1.3 to uncountably many H. Spectrum and eigenfunctions of G can be split in infinitely many ways to obtain H.

To satisfy time-reversal (see section 1.2), we split complex eigenvectors into real and imaginary parts, and allocate them to eigenvalues meeting the distance criterion of formula 1.3, thus achieving a H with non-degenerate eigenvalues.

b) As the set of pure inputs is countable, the spectral measure of all the orbits can be allocated to disjoint, finite intervals. As each orbit has bounded spectrum, it has also finite energy.          q.e.d.

### 1.4.4 Physical Complexity

The straight-forward way to compare the computational strength of QTMs and physical implementations would be by the use of real time complexity, taking the common step time for QTMs and continuous, physical time for physical implementations. However, the definition of physical implementations includes the machine cycle time T, which can be arbitrarily chosen. By stretching the energy scale, the time scale is squeezed reciprocally. We therefore seek a more robust and less arbitrary measure.

The action $\langle H \rangle \frac{T}{\hbar}$ would be invariant under dilatations, but still a translation could make it zero for an arbitrary $q$ (not necessarily an eigenvector of H) and negative for some subspace. Our construction in 1.4.3 could be modified to yield zero for all $q$, even with H of bounded spectrum! $\langle -\Delta \rangle \frac{T}{\hbar}$ would not have this deficiency, but not suitably reflect the total action and the effect of potentials. We end up at

Definition 1.3

Let $q_t = e^{-i\frac{Ht}{\hbar}} q_0$. Then the *physical complexity* of the physical transform $q_0 \xrightarrow{H} q_t$ is

$$C(q_t, q_0 | H) = \frac{t}{\hbar} \langle |H| \rangle,$$

where we define

$$\langle |H| \rangle = \int_\lambda |\lambda| d\mu_q(\lambda),.$$

This measure is dimensionless, independent from scale, non-negative, and zero only if q is an eigenvector of H with eigenvalue 0, i.e. invariant under U. It is elementary to see that $\langle |H| \rangle$ is a constant of motion, i.e. constant on any orbit.

The following proposition establishes a lower bound by $L^2$-distance.

Proposition 1.4

For any physical state $q$ and Hamiltonian H, physical complexity is lower-bounded by half the square $L^2$ space distance

$$\|q_t - q_0\|_2^2 \leq 2C(q_t, q_0 | H).$$

Particularly, evolution into an orthogonal state at least takes physical complexity 1.

Proof:
By the Schrödinger equation, and recalling the time constancy of the norm on orbits,

$$\frac{d}{dt}\|q_t - q\|_2^2 = \left(-i\frac{H}{\hbar}q_t, q\right) + \left(q, -i\frac{H}{\hbar}q_t\right)$$

$$= 2\int_\lambda \frac{\lambda}{\hbar}\sin\frac{\lambda t}{\hbar}d\mu_q \quad \lambda$$

$$\leq \frac{2}{\hbar}\langle H\rangle$$

$$\leq \frac{2}{\hbar}\langle |H|\rangle < \infty$$

$\langle H\rangle$ is finite as we require physical states to have finite energy (see section 1.3). This implies finiteness of $\langle |H|\rangle$ by elementary measure theory. The claimed result follows immediately from definition 1.3.          q.e.d.

One physical complexity unit corresponds to the evolution of a state into an orthogonal state, exactly like one computational step of a DTM. When comparing a DTM M with the physical implementation P of an equivalent QTM M', then by proposition 1.4 one DTM step will take at least physical complexity 1, as DTMs evolve into orthogonal states.

Definition 1.4:
    a) The *time complexity* of a computation performed by a HA is the physical complexity of the physical transform of the initial to the final state.
    b) A physical implementation is *efficient*, if it performs computations with physical complexity of the same order as the QTM implemented.

When necessary, we will denote this by HA time complexity or physical complexity, to distinguish it from Turing model time complexity. Computational time will refer to (Turing or HA) time complexity, if a distinction from real time is needed.

Notice that efficiency implies that all pure inputs map to physical states with finite energy.

We relax now and in the following the definition of physical implementations given in section 1.4.1 by allowing T to depend on the problem size n. Proposition 1.3 can be strengthened to claim the existence of efficient physical implementations.

Proposition 1.5
Let P be an efficient implementation of some QTM M. Then there is a constant $c > 0$ such that such that for any $q$, $d\mu_q^\wedge(t)$ has no more than $c$ zeros in the interval $[0, T]$.
Proof:
By proposition 1.4, the number of zeros in the interval $[0, T]$ is a lower bound for twice the physical complexity of the physical transform $q_0 \xrightarrow{H} q_T$. By the semi-group property of the evolution operator, the physical transform $q_t \xrightarrow{H} q_{t+T}$ has the same physical complexity. The time complexity of a computation is thus proportional to this number. By efficiency, it must be upper bounded by $c$, uniformly for all inputs.          q.e.d.

As this proposition shows, the gauge condition rules out approaches such as [Co98] based on halving $T$ from step to step, as the HA time complexity of the process would be infinite after finite real time $T$.

## 2    Computational Results

We define instant computation, $\alpha-$ waiting DTMs as well as instant-computing and standard implementations of DTMs, and prove important properties, particularly the existence of good implementations. This sets the basis for the computational results of theorem 1.

### 2.1    Instant Computation

Definition 2.1

a) $P_X$ is the projection operator on Hilbert space $X$.

b) Any computation of a function $f(x)$, maintains a result variable $r = (z, v)$, where $z = 0$ in result states (halting states), $z \neq 0$ otherwise, and $r = f(x)$ whenever $z = 0$.

c) For any computation, predicate $\Pi(r)$ is true iff $r$ is the true result of that computation, i.e. for halting computations, $r = (0, f(x))$, for non-terminating computations $z \neq 0$. $\Pi(q)$ is true iff the state $q$ contains the true result.

d) $\Theta = Span(\{q_{kT} | k \in \mathbb{Z} \})$ is the span of all computational states on orbit $q_t$.

e) $\Xi = Span(\{q_{kT} | k \in \mathbb{Z} \wedge \Pi(q_{kT})\})$ is the span of all computational states on orbit $q_t$ containing the true result.

f) An *instant computation configuration* $C = (P, q_0, \tau, O)$ is a HA P with Hamiltonian H, $0 < \tau < T$ and initial state $q_0$, such that the $q_{kT}$ are mutually orthogonal ($0 \leq k < p$), where $p \leq \infty$ is the period of orbit $q_t$, furthermore an observable O which has eigenspace $\Omega$ at eigenvalue 1 such that $\Theta \subset \Omega$. Measurement of O thus yields the value 1 for computational states.

g) $\nu_q = \sum_{k=0}^{p} \left| d\mu_q^\wedge(kT + \tau) \right|^2 = \|P_\Theta q_\tau\|_2^2$ is the probability of measuring a computational state at time $\tau$.

h) $\nu_{q,c} = \dfrac{\nu_q}{\|P_\Omega q_\tau\|_2^2}$ is the probability of measuring a computational state at time $\tau$, conditional under $O = 1$ at time $\tau$.

i) $\pi_{q,c} = \dfrac{\|P_\Xi q_\tau\|_2^2}{\|P_\Omega q_\tau\|_2^2}$ is the probability of obtaining the true computational result at time $\tau$, conditional under $O = 1$ at time $\tau$.

j) $C$ is $\delta-yielding$ iff $\nu_q \geq \delta > 0$.

k) $C$ is $\varepsilon-valid$, iff $\pi_{q,c} > \varepsilon$.

l) Let M be a DTM. An implementation P of M is $\delta_n - \varepsilon_n -instant\text{-}computing$, iff for problem size $n$, $0 < \tau_n < T_n$ and any input $q_0$, $C = \left( P, q_0, \tau, O \right)$ is a $\delta_n$–yielding, $\varepsilon_n$–valid instant computation configuration.

m) P is *validable*, iff $\Pi(r)$ is computable.

n) P is *good*, iff P is efficient and $\delta_n > \delta > 0$, $\varepsilon_n > \varepsilon > 0$ uniformly, where $\varepsilon > \dfrac{1}{2}$ for non-validable implementations.

<u>Proposition 2.1: Instant Computation</u>
Let $C$ be $\delta$ – yielding and $\varepsilon$ – valid, where $0 < \delta \leq 1$.

### a) Error-free procedure

Let $C$ be validable, $0 < \varepsilon \leq 1$. Then the following physical procedure yields in expected time

$$O\left( \frac{\tau + \tau_0 + \tau_1}{\delta} + \frac{\tau_2 + \tau_3}{\varepsilon} \right)$$ a result $r$ valid with certainty, where $\tau_0$ is preparation time of $q_0$, $\tau_1$

the measurement time of O, $\tau_2$ the measurement time of $r$ and $\tau_3$ the evaluation time of $\Pi$.

1. Prepare $q_0$.
2. Let H act on $q_0$ for time $\tau$.
3. Measure O.
4. If the value does not equal 1, then go to step 1.
5. Measure the result $r$.
6. If $\Pi(r)$ is false, then go to step 1.
7. Output the result $r$.

### b) Error-bounded procedure

Let $\varepsilon > \frac{1}{2}$. Then the following physical procedure yields in expected time $O\left( \frac{\tau + \tau_0 + \tau_1}{\delta} + \tau_2 \right)$ a result $r$ such that $\Pi(r)$ is true with probability $\geq \varepsilon$. With standard majority techniques the error probability can be made arbitrarily small.

1. Prepare $q_0$.
2. Let H act on $q_0$ for time $\tau$.
3. Measure O.
4. If the value does not equal 1, then go to step 1.
5. Measure $r$.
6. Output the result $r$.

Proof:
a) Step 3 yields value 1 with probability $\left\| P_\Omega q_\tau \right\|_2^2$, and step 6 is true with probability $\pi_{q,c}$. In the mean, steps 1-4 are executed $O\left( v_q^{-1} \right)$ times, steps 5-6 $O\left( \pi_{q,c}^{-1} \right)$ times. This implies the time estimates. At step 7, $\Pi(r)$ is always true. This proves part a).

b) O is assumed to be (classically) correlated with the states $q_{kT}$ such that $v_{q,c} \geq \pi_{q,c} > \varepsilon > \frac{1}{2}$. Step 3 yields value 1 with probability $\left\| P_\Omega q_\tau \right\|_2^2 > \delta$. At step 6, $O = 1$ with certainty, thus $\Pi(r)$ is true with probability $\pi_{q,c} > \frac{1}{2}$. This proves part b).

q.e.d.

## 2.2    $\alpha$-Waiting DTMs

<u>Definition 2.2</u>
Let M be a DTM. M is $\alpha-$waiting, iff for any input c in M's domain of definition, M is periodic on c with period $p < \infty$, and after finding the result in step $k$, M remains in result states and maintains the valid result on the working tape for at least $\alpha p$ steps $(0 < \alpha < 1)$. Notice that if M does not halt on c, then $z \neq 0$, thus $r$ is always valid.

<u>Proposition 2.2</u>
For every DTM M and $0 < \alpha < 1$ there is an equivalent, $\alpha-$waiting DTM M' with the same order of Turing-computational time complexity.

Proof:
M' is easily constructed from M. We first assume w.l.o.g. that M operates reversibly (see [Be73, 82, 93, 95] )

1. Execute M, whereas counting the steps. I.e. after each step increase a step count $s$ before executing the next step.
2. When a result state is reached, then set $s' = \dfrac{\alpha}{1-\alpha} s$, and count $s'$ down to zero, whereas inserting an additional waiting state after every decrease operation.
3. Undo step 2 (see Bennett for reversible undoing).
4. Undo step 1.
5. Go to step 1.

The period of M' is a constant multiple of the period of M, because increase and decrease operations on binary numbers take average two bit operations.                    q.e.d.

## 2.3     Standard Implementations

<u>Definition 2.3</u>
Let M be a DTM computing a function $f(x)$. Then its standard implementation is defined as follows: M is transformed into an equivalent $\alpha-$waiting DTM M'. M' is implemented into a QTM M''. M'' is implemented into an instant-computing HA P.

<u>Proposition 2.3</u>
Let M be a DTM computing a function $f(x)$, P an efficient, instant-computing standard implementation of M. If M is total and P validable, then the error-free procedure according proposition 2.1a) is applicable. Otherwise, if $\varepsilon > \dfrac{1}{2}$, then the error-bounded procedure according proposition 2.1b) is applicable.

Proof:
If M is total, M always terminates, and as P is validable, the error-free instant computation procedure will eventually terminate. At step 7, $r$ is valid.

If M is total or partial and $\varepsilon > \dfrac{1}{2}$, then the error-bounded instant computation procedure will eventually find $O = 1$ in step 3 and thus reach step 6 with a value of $r$ which is valid with probability $> \varepsilon > \dfrac{1}{2}$. As there is a unique valid value, repetition of this procedure and taking the majority result can be used to obtain the valid result with arbitrarily small error probability.

Recalling the definitions of $r$ and $\Pi$, one verifies that this holds for terminating as well as for non-terminating computations.                              q.e.d.

## 2.4    Good Implementations

In this section we prove the (mathematical) existence of good implementation. In the remaining part of this paper we will always take $\tau = \dfrac{T}{2}$ .

<u>Definition 2.4</u>
An    implementation    is    *minimal*    iff    for    each    pure    input $q_0 \colon Span(\{q_t | t \in \mathbb{R}\}) = Span(\{q_{kT} | k \in \mathbb{Z}\})$.

<u>Proposition 2.4</u>
For any DTM M, mapping $\varphi$, define basis B and O as in proposition 1.3. Then the following holds:
a) M has infinitely many good standard implementations.
b) Infinitely many of these implementations have Hamiltonians H such that, for any periodic orbit, the restrictions of H to that orbit is computable from B.
c) Unless M is total, neither H nor the evolution operator $U_{T/2}$ is enumerable from B.

Proof:
a) We explicitly construct minimal implementations and prove that they are good.

First consider a periodic orbit with initial state $q_0$. Define the spectrum of $\dfrac{HT}{\hbar} \bmod 4\pi$ such that

(2.4.1) $$\lambda_k = 2\pi \left( \frac{k}{p} + (k \bmod 2) \right) \ (k = 0, \ldots, p-1),$$

and the eigenstates $f_k$ by formula 1.4, such that

$$q_t = \frac{1}{\sqrt{p}} \sum_{k=0}^{p-1} e^{-2\pi i \left( \frac{k}{p} + k \bmod 2 \right) t} f_k \ .$$

Noticing that by definition 2.2, p is even, we obtain

$$d\mu_q^{\hat{}}\left( \left( j - \frac{1}{2} \right) T \right) = \frac{1}{p} \sum_{k=0}^{p-1} e^{2\pi i \left( \frac{k}{p} + k \bmod 2 \right) \left( j - \frac{1}{2} \right)}$$

$$= \frac{1}{p} \sum_{k=0}^{p-1} -1^{\ k} \ e^{2\pi i \frac{k}{p} \left( j - \frac{1}{2} \right)}$$

$$= \frac{1}{p} \frac{1 - e^{2\pi i \left( j - \frac{1}{2} \right)}}{1 + e^{\frac{2\pi i}{p} \left( j - \frac{1}{2} \right)}}$$

$$= \frac{1}{p} e^{\frac{\pi i}{p}\left(j-\frac{1}{2}\right)} \frac{2}{e^{\frac{\pi i}{p}\left(j-\frac{1}{2}\right)} + e^{-\frac{\pi i}{p}\left(j-\frac{1}{2}\right)}}$$

$$= \frac{e^{\frac{\pi i}{p}\left(j-\frac{1}{2}\right)}}{p \cos \frac{\pi}{p}\left(j-\frac{1}{2}\right)},$$

whose absolute value has maximum $\approx \frac{2}{\pi}$ at $j = \frac{p}{2}$ and, when squared, sums up to 1 due to minimality. Due to definition 2.2, 2.3 and 2.4,

$$\nu_q = 2 \sum_{k=(1-\alpha)p}^{p} \left| \frac{1}{p \cos \frac{\pi}{p}\left(k-\frac{1}{2}\right)} \right|^2$$

$$= 1 - 2 \sum_{k=0}^{(1-\alpha)(p-1)} \left| \frac{1}{p \cos \frac{\pi}{p}\left(k-\frac{1}{2}\right)} \right|^2$$

$$\geq 1 - \frac{2(1-\alpha)}{p \cos \pi (1-\alpha)}$$

$$\to 1 \quad (\alpha \to 1)$$

Next consider an aperiodic orbit with initial state $q_0$. Define the spectrum of $\frac{HT}{\hbar} \bmod 2\pi$ such that

$$d\mu_q(\lambda) = \frac{d\lambda}{2\pi} \quad (0 < \lambda < 2\pi),$$

and the spectral transform maps $q_t$ to $e^{-i\lambda \frac{t}{T}} \in L^2\left(d\mu_q\right)$, thus

$$d\hat{\mu_q}\left(\left(k-\frac{1}{2}\right)T\right) = \frac{1}{2\pi} \int_0^{2\pi} e^{-i\left(k-\frac{1}{2}\right)\lambda} d\lambda$$

$$= \frac{1}{2\pi i} \left| \frac{-e^{-i\left(k-\frac{1}{2}\right)\lambda}}{k-\frac{1}{2}} \right|_0^{2\pi}$$

$$= \frac{-1}{\pi i \left(k-\frac{1}{2}\right)}$$

Due to definition 2.4,

$$v_q = \frac{1}{\pi^2} \sum_{-\infty}^{+\infty} \frac{1}{\left(k - \frac{1}{2}\right)^2}$$

$$= \frac{8}{\pi^2} \sum_{k \geq 0} \frac{1}{(2k+1)^2}$$

$$= 1$$

by a result from Euler (see also [He88], formula 4.9-4). $v_q = 1$ is a direct consequence of the minimality of the implementation. The modular constraint leaves infinite degrees of freedom for the allocation of the spectrum. This proves part a).

b) Matching the above construction with proposition 1.3 and definition 1.2, one easily verifies that the restriction of H is computable from the eigenfunctions and these from B, proving this part.

c) For non-halting computations, eigenfunctions do not exist, and convergence of finite approximations is only weak.                                   q.e.d.

## 2.5    Computational Power

Definition 2.5
a) The class *HRP (Hamiltonian Random Polynomial)* consists of those problems solvable by a HA in random polynomial time without error.
b) The class *HBPL (Hamiltonian Bounded Probabilistic Linear)* consists of those problems solvable by a HA in random linear time with bounded error.

Theorem 1
a) $NEXPPOLY \subseteq HRP$
b) For any (partial or total) recursive function $f$ define $g$ by

$$g(n) = \begin{cases} (1, f(n)) & (n \in D(f)) \\ 0 & (n \notin D(f)) \end{cases}$$

Then $g \in HBPL$.

Proof:
a) By theorem 11.4 of [DK00], $NEXPOLY = PCP(poly, poly)$, the class of probabilistically checkable proof systems with polynomial number of random bits and oracle queries. Following the construction of the non-adaptive oracle PTM in [DK00], page 395, we define a DTM M which, on input $(n, y)$ returns $(f(n), z)$, where $y$ is a polynomial length random string and $z$ a polynomial length bit string constituting part of a proof for $n \xrightarrow{f} f(n)$.

M computes $z$ by simulating a polynomial number of oracle queries. This is possible, as $f$ is total. M acts as prover and as well selects a polynomial size part from the exponential size proof, according to the random input. Whereas in proof systems this selection would be the verifier's task, in our setting it can well be done by M. Alternatively, M could create a superposition of the exponential set of polynomial size proofs, appending a sequential number as part of the result. Measurement of this number part would yield one proof instance perfectly randomly.

A good instant-computing implementation of M will positively validate $z$ with one-sided error $\leq \dfrac{1}{4}$ , according to definition 11.2 of [DK00] and proposition 2.3. Such a good implementation exists by proposition 2.4. The number of iterations is $O_p(poly)$ and the execution time of each step is polynomially bounded. HA time complexity is proportional to real time complexity, as the Hamiltonian has bounded spectrum. q.e.d.

b) Let M be a DTM computing the (total or partial) function $f$ , as described in definition 2.3. By proposition 2.4 a good instant-computing implementation of M exists and yields the result with arbitrarily small error by proposition 2.3. The number of iterations is $O_p(1)$ and the execution time of each step $O(n)$. HA time complexity is proportional to real time complexity, as the Hamiltonian has bounded spectrum. q.e.d.

In the view of theorem 1b and theorem 3 below, one may question the need for theorem 1a. The rationale is the fact, that $\varepsilon^{-1} \in O_p(poly)$ is sufficient for the error-free instant computing procedure. Therefore, the range of implementations realizing theorem 1a is potentially larger.

It easy to see how good implementations can be relaxed to yield implementations which are weaker but still useful.

## 3    Realization Issues

We derive an interesting result on implementation complexity (theorem 2), prove a fundamental theorem on the physical realizability of instant computation (theorem 3), and introduce Schrödinger Automata (SA), pointing out difficulties arising in physical implementations in such automata.

### 3.1    Implementation Complexity

One may ask what amount of resources (energy, matter and space) is required for a QTM implementation. Obviously, a classical physical implementation of the restriction of a DTM to problems up to some size $n$ requires matter, space and energy proportional to the DTM's space complexity, $Space(n)$, mainly for the working tape, plus dissipated energy proportional to the time complexity, $Time(n)$. An unrestricted implementation needs infinite resources.

The following theorem shows that the HA on which theorem 1 relies can be implemented with finite resources. Though the construction is purely mathematical and for various reasons not physically realizable, this is an important result showing that not only finite automata but even full-fledged QTMs do not a priori require infinite capabilities.

Theorem 2
Every total DTM M possesses an efficient implementation with the following properties
   a) P is constructible from B.
   b) P is good.
   c) The Hamiltonian has non-degenerate eigenvalues.
   d) P is implemented in the subspace $L^2(\mathbb{R})$ of the single particle state space.
   e) Bounded space is required to hold any problem instance, for unrestricted problem size $n$.
   f) Bounded energy is required to represent any problem instance, for unrestricted problem size $n$.
   g) P is capable to solve any problem instance, for unrestricted problem size $n$.

Proof: Let P be a minimal implementation as defined in section 2.4. There are $2^n$ instances of problems of size n. Due to the totality of M, we may w.l.o.g. assume that the period is $p_n \leq 2^{\nu_n}$ for some strictly increasing $\nu_n$. We construct an implementation P into $L^2(\mathbb{R})$ (i.e. a subspace of the single-particle state space) by an economical assignment of eigenfunctions to eigenvalues, such that different problem instances have disjoint spectrum. Thus all instances of problem size n together occupy $2^{n+\nu_n}$ eigenvalues. We choose these eigenvalues such that eigenvalue $\lambda_k \ (0 \leq k < 2^{\nu_n})$ of instance $m \ (0 \leq m < 2^n)$ satisfies $\frac{\lambda_k}{2\pi} = m2^{-n-\nu_n} + k2^{-\nu_n} \bmod 1$, and $\frac{\lambda_k}{2\pi} \in [k \bmod 2, k+1 \bmod 2)$, in the same way as in proposition 2.5..

Now we make the following induction hypothesis: When the spectrum to all problems of size $\leq n$ is assigned, then interval $[2\pi k, 2\pi(k+1)) \ (0 \leq k < n)$ contains $2^{n-k+\nu_n}$ equidistant eigenvalues. We start the induction by assigning $\lambda_0 = 0$ to the problem of size $n = 0$ and defining $\nu_0 = 0$. Having assigned the spectrum to all problems of size $< n$, we assign the spectrum to problems of size $n$ as follows: We equidistantly fill $2^{n-k+\nu_n} - 2^{n-k-1+\nu_{n-1}}$ points into interval $[2\pi k, 2\pi(k+1)) \ (0 \leq k < n)$ and $2^{\nu_n}$ points into interval $[2\pi n, 2\pi(n+1))$. The induction hypothesis is obviously satisfied by this construction. All $2^{n+\nu_n}$ points are assigned, as the sum telescopes.

Now a), c) and g) above are satisfied by construction. f) follows from the estimate $\sum_{k=0}^{n} \pi k 2^{-k} \leq \pi \left. \dfrac{d}{dx} \dfrac{1}{1-x} \right|_{x=\frac{1}{2}} = 4\pi$, which bounds the energy of any problem instance of any size.

d) and e) are easily satisfied by choosing B accordingly. One verifies b) by following the proof of proposition 2.4 and considering

$$\nu_n = 2^{-\nu_n} \sum_j \left| \sum_k -1^k \, e^{2\pi i \, m 2^{-n-\nu_n} + k 2^{-\nu_n} + 2\zeta_k \left(j-\frac{1}{2}\right)} \right|^2$$

$$= 2^{-\nu_n} \sum_j \left| \sum_k -1^k \, e^{2\pi i \, k 2^{-\nu_n} + 2\zeta_k \left(j-\frac{1}{2}\right)} \right|^2$$

$$\geq \frac{3}{4} 2^{-\nu_n} \sum_j \left| \sum_k -1^k \, e^{2\pi i \, k 2^{-\nu_n} \left(j-\frac{1}{2}\right)} \right|^2$$

$$\approx \frac{3}{\pi^2}$$

for some integer-valued $\zeta_n \geq 0$ in $\frac{7}{8}$ of all cases.                    q.e.d.

Admitting degenerate eigenvalues, any DTM can be implemented into a HA with bounded spectrum. If in addition the DTM is total, then its state will be bounded. If it is partial, then the state of non-halting computations dissipates to infinity, by the RAGE theorem [RS80].

This construction is purely mathematical. The above Hamiltonian has dense point spectrum on the positive real half-axis. Physical particle states with arbitrarily high energy levels occupied are unstable. However, real physical systems may approximately realize this construction.

## 3.2    Realizability

Whereas proposition 2.4 proves the (mathematical) existence of uncountably many good, minimal implementations, it still leaves open the physical realizability of this construction. We provide a partial answer by showing that randomly chosen HA implementations of $\alpha-$ waiting DTMs are likely to be good.

In other words: The capability to realize general QTM implementations is likely to imply the capability to realize good QTM implementations. Instant computing requires no special technology beyond general QTM implementation technology. However, QTMs cannot be realized today, and nobody knows if they ever will be.

### 3.2.1   Statistical Properties

We are now going to analyze the statistical properties of the class of all efficient HA implementations of (total and partial) $\alpha-$ waiting DTMs.

3.2.1.1 Sampling Distribution of Efficient Implementations

Recall that any periodic orbit with period $p$ has pure point spectrum, its eigenvalues satisfy $\dfrac{\lambda_k T}{\hbar} = 0 \bmod \dfrac{2\pi}{p}$ and its spectral measure

$$\mu_q \left( \left\{ \lambda_k \Big| \frac{\lambda_k T}{\hbar} = \frac{2\pi j}{p} \bmod 2\pi \right\} \right) = \frac{1}{p} \quad (j = 0, \dots, p-1),$$

as otherwise $d\mu_q^\wedge(kT)$ would not equal $\delta_{k \bmod p}$. Define

$$x_{j,s} = \mu_q\left(\left\{\lambda_k \Big| \frac{\lambda_k T}{\hbar} = 2\pi\left(s + \frac{j}{p}\right) \bmod 4\pi\right\}\right) \quad (j = 0, \ldots, p-1, s = 0,1).$$

Any implementation must satisfy $x_{j,0} + x_{j,1} = \frac{1}{p}$, whereas $y_j = x_{j,0} - x_{j,1}$ may take an arbitrary value in the interval $\left[-\frac{1}{p}, +\frac{1}{p}\right]$. An engineer constructing a HA does not need to control $y_j$.

If we take a random sample from all possible efficient designs, then the $y_j$ will have some distribution $F_p(y_j)$. For symmetry reasons, it will be the same for all j, i.e. they are i.i.d. with distribution $F_p(y_j) = F_p(y)$. Furthermore, by efficiency (see proposition 1.5), there is no fundamental difference between different orbits of the same period, thus the above distribution is valid for all of them. Again by efficiency, the only difference between orbits of different periods is the period length, not the underlying physical process or implementation technique, as we are sampling over all possible efficient designs. Therefore $F_p(y) = F(py) \left(-\frac{1}{p} \leq y \leq \frac{1}{p}\right)$ for some underlying distribution $F(z)$ on $[-1, +1]$. Finally, as HA with Hamiltonians $H$ and $H + 2\pi$ are computationally equivalent, but have $y_j$ of opposite sign, the probability density $f$ is an even function, as a matter of fact bi-modal. We conclude:

Lemma 3.1

The $y_j$ $(j = 0, \ldots, p-1)$ are i.i.d. random variables with density $pf(py)$ $\left(-\frac{1}{p} \leq y \leq \frac{1}{p}\right)$, where $f$ is even. The odd moments are zero, and the even moments $m_{2n}p^{-2n}$, where $m_n$ is the n-th moment of $f$.

3.2.1.2 Success Probability: Periodic Orbits

We calculate now the success probability $v_q$ and its moments in the periodic case. First, notice that

$$d\mu_q^\wedge\left(\left(j - \frac{1}{2}\right)T\right) = \sum_{k=0}^{\infty} |\alpha_k|^2 e^{-i\lambda_k\left(j - \frac{1}{2}\right)}$$

$$= \sum_{k=0}^{p-1} y_k e^{-2\pi i \frac{k}{p}\left(j - \frac{1}{2}\right)}.$$

Due to the independence of the $y_k$, and because $m_1 = 0$,

$$E\left(\left|d\mu_q^\wedge\left(\left(j - \frac{1}{2}\right)T\right)\right|^2\right) = E\left(\left|\sum_{k=0}^{p-1} y_k e^{-2\pi i \frac{k}{p}\left(j - \frac{1}{2}\right)}\right|^2\right)$$

$$= E\left(\sum_{k=0}^{p-1} y_k^2\right) = pm_2 p^{-2}$$

$$= p^{-1}m_2$$

This holds independently for all $j = 0, \ldots, p-1$, implying $E(j^n) = O(p^n)$, which is in itself interesting result. By definition 2.1.g and the linearity of the expectation operator we immediately conclude

(3.1)
$$E(v_q) = \alpha m_2.$$

As the waiting time of any $\alpha - $ waiting DTM can be made arbitrarily long by a suitable waiting rule, yielding correspondingly long periods $p$, we set

(3.2)
$$\alpha = 1 - p^{-1/2}.$$

This leads to

$$\begin{aligned}
v_q &= \sum_{j=p^{\frac{1-\alpha}{2}}}^{p^{\frac{1+\alpha}{2}}} \sum_{k=0}^{p-1} \sum_{l=0}^{p-1} y_k y_l e^{-2\pi i \frac{k-l}{p}\left(j-\frac{1}{2}\right)} \\
&= \sum_{k=0}^{p-1} \alpha p y_k^2 + \sum_{l \neq k} y_k y_l e^{-\pi i (k-l)\left(1-\alpha-\frac{2}{p}\right)} \frac{1 - e^{-2\pi i (k-l)\left(\alpha+\frac{1}{p}\right)}}{1 - e^{-2\pi i \frac{k-l}{p}}} \\
&= \sum_{k=0}^{p-1} \alpha p y_k^2 + \sum_{l \neq k} y_k y_l e^{-\pi i (k-l)\left(1-\frac{2}{p}\right)} \frac{e^{+\pi i (k-l)\left(\alpha+\frac{1}{p}\right)} - e^{-\pi i (k-l)\left(\alpha+\frac{1}{p}\right)}}{e^{+\pi i \frac{k-l}{p}} - e^{-\pi i \frac{k-l}{p}}} \\
&= \sum_{k=0}^{p-1} \alpha p y_k^2 + \sum_{l \neq k} y_k y_l e^{2\pi i \frac{k-l}{p}} \frac{\sin \pi \frac{k-l}{\sqrt{p}}\left(1 - \frac{1}{\sqrt{p}}\right)}{\sin \pi \frac{k-l}{p}} \\
&\approx \sum_{k=0}^{p-1} \alpha p y_k^2 + 2 \sum_{l < k} y_k y_l \cos\left(2\pi \frac{k-l}{p}\right) \frac{\sin \pi \frac{k-l}{\sqrt{p}}}{\sin \pi \frac{k-l}{p}}
\end{aligned}$$

Formula 3.1 is directly verified using independence and the fact that all odd moments are zero. For the second moment one finds, again dropping the odd moments,

$$\begin{aligned}
E(v_q^2) &- E^2(v_q) = \\
&= E\left(\left(\sum_{k=0}^{p-1} \alpha p y_k^2 + 2 \sum_{l < k} y_k y_l \cos\left(2\pi \frac{k-l}{p}\right) \frac{\sin \pi \frac{k-l}{\sqrt{p}}}{\sin \pi \frac{k-l}{p}}\right)^2\right) - \alpha^2 m_2^2 \\
&= -\alpha^2 m_2^2 + \alpha^2 p^2 (p m_4 + p^2 m_2^2) p^{-4} + \sum_{k=0}^{p-1} 4 m_2^2 p^{-4} \sum_{l < k} y_k y_l \cos^2\left(2\pi \frac{k-l}{p}\right) \frac{\sin^2 \pi \frac{k-l}{\sqrt{p}}}{\sin^2 \pi \frac{k-l}{p}} \\
&= \alpha^2 p^{-1} (m_4 - m_2^2) + o(p^{-1} m_2^2)
\end{aligned}$$

The $o(p^{-1})$ term is obtained by finding $o(p^2)$ upper bounds for the r.h.s. of

$$\sum_{k=0}^{p-1} 2 \sum_{l < k} \cos^2\left(2\pi \frac{k-l}{p}\right) \frac{\sin^2 \pi \frac{k-l}{\sqrt{p}}}{\sin^2 \pi \frac{k-l}{p}} \leq p \sum_{k=0}^{p-1} \frac{\sin^2 \pi \frac{k}{\sqrt{p}}}{\sin^2 \pi \frac{k}{p}} \leq$$

For the disjoint sets $k \, \epsilon \, o\left(p^{1/4}\right)$, $k \, \epsilon \, o\left(p^{1/2}\right) - o\left(p^{1/4}\right)$, $k \, \epsilon \, o(p) - o\left(p^{1/2}\right)$ and $k \, \epsilon \, o(p)$.

By the Chebychev inequality

(3.3)
$$\nu_q = \alpha m_2 + O_p\left(p^{-1/2}\right).$$

### 3.2.1.3 Aperiodic Orbits

We adapt the analysis of the previous section to the continuous case, using the distributional calculus [Ru73]. First we see that

$$d\mu_q^{\wedge}\left(\left(j - \frac{1}{2}\right)T\right) = \int_{-\infty}^{+\infty} e^{-i\lambda\left(j-\frac{1}{2}\right)} d\mu_q(\lambda)$$
$$= \frac{1}{2\pi}\int_0^{2\pi} e^{-i\lambda\left(j-\frac{1}{2}\right)} y(\lambda) d\lambda$$

where $y(\lambda) = x_0(\lambda) - x_1(\lambda), 0 \leq x_0(\lambda) \leq 1, \ x_0(\lambda) + x_1(\lambda) = 1$. Similar arguments as in section 3.2.1 imply that $y(\lambda)$ has even density $g(y)$ independent from $\lambda$, with moments $m'_n$, and the stochastic independence for $\lambda \neq \lambda'$ is expressed by

$$E\left(y^r(\lambda)y^s(\lambda')\right) = E\left(y^r(\lambda)\right)E\left(y^s(\lambda)\right) \ (\lambda \neq \lambda').$$

This leads to

$$E\left(\left|d\mu_q^{\wedge}\left(\left(j - \frac{1}{2}\right)T\right)\right|^2\right) = \frac{1}{4\pi^2}\int_0^{2\pi}\int_0^{2\pi} E\left(y(\lambda)y(\lambda')\right)e^{-i(\lambda-\lambda')\left(j-\frac{1}{2}\right)} d\lambda d\lambda'$$
$$= 0$$

in line with the discrete case. The value of $\nu_q$ is

$$\nu_q = \sum_{j=-\infty}^{+\infty} \frac{1}{4\pi^2}\int_0^{2\pi}\int_0^{2\pi} y(\lambda)y(\lambda')e^{-i(\lambda-\lambda')\left(j-\frac{1}{2}\right)} d\lambda d\lambda'$$
$$= \lim_{m\to\infty} \frac{1}{4\pi^2}\int_0^{2\pi}\int_0^{2\pi} y(\lambda)y(\lambda')e^{-i\frac{\lambda-\lambda'}{2}}\frac{\sin m(\lambda - \lambda')}{\sin\frac{\lambda - \lambda'}{2}} d\lambda d\lambda'$$

where taking the principal value is justified by the fact that $\left|\nu_q\right| \leq 1$. For further evaluation, we first notice that the functions $h_m(x) = \frac{\sin mx}{\sin\frac{x}{2}}$ $(-\pi \leq x \leq \pi)$ have a removable singularity at the origin and asymptotically vanishing integral for for $x \in \Omega(m^{-1})$, whereas for $x \in o(m^{-1})$ the approximations

$$h_m(x) = \frac{m}{2}\frac{\sin mx}{\frac{mx}{2}} + o(1)$$
$$= m\cos m\frac{x}{2} + o(1)$$

hold. Approximating $\cos x$ in $\left[-\frac{\pi}{2}, +\frac{\pi}{2}\right]$ by test functions and applying theorem 6.32 from [Ru73] yields a representation of Dirac's delta functional, $\lim_{m\to\infty} m\cos mx = 2\delta_x$, such that

$$\lim_{m\to\infty} \frac{1}{2\pi} \int_{-\pi}^{+\pi} f(x)m\cos m\frac{x}{2}\,dx = 2f(0).$$

The above arguments and the properties of $h_m$ show that

$$\lim_{m\to\infty} e^{-i\frac{\lambda-\lambda'}{2}}\, h_m(x) = \delta_{\lambda-\lambda'} + \delta_{\lambda-\lambda'+\pi \bmod 2\pi}.$$

One obtains

$$v_q = \frac{1}{4\pi^2} \int_0^{2\pi}\int_0^{2\pi} y(\lambda)y(\lambda')\left(\delta_{\lambda-\lambda'} + \delta_{\lambda-\lambda'+\pi \bmod 2\pi}\right)d\lambda d\lambda'$$
$$= \frac{1}{2\pi}\int_0^{2\pi}\left(y^2(\lambda) + y(\lambda)y(\lambda+\pi \bmod 2\pi)\right)d\lambda,$$

which immediately yields
$$E(v_q) = m_2',$$

in line with the discrete case. Due to independence, the second moment is found to be

$$E(v_q^2) = E\left(\left(\frac{1}{2\pi}\int_0^{2\pi}\left(y^2(\lambda) + y(\lambda)y(\lambda+\pi \bmod 2\pi)\right)d\lambda\right)^2\right)$$
$$= E(v_q)$$

Thus $v_q$ has zero variance, again in conformance with the discrete case. The analysis is summarized by the following theorem:

### Theorem 3
Let M be an $\alpha-$waiting DTM such that $\alpha = 1 - p^{-1/2}$ for each periodic orbit. Let P be obtained by random sampling from the set of all efficient standard implementations of M. Then the following holds:

1. Any instant computation configuration C is $\delta-$yielding for $\delta$ as defined below.
2. If C is periodic then $\delta < m_2 - \frac{\Delta}{c\sqrt{p}}$ with probability greater than $1 - \frac{1}{\Delta^2}$, for some $c > 0$.
3. For arbitrary $\wedge\varepsilon > 0 : \delta > m_2 - \varepsilon$ with probability $1 - O(p^{-1})\,(p\to\infty)$.
4. If C is aperiodic then $\delta = m_2'$ with probability 1.

Proof: Immediately from the previous derivations and definitions 2.1, 2.2 and 2.3.

q.e.d.

#### 3.2.2  Validity

Definition 2.1 implies $v_q \le \pi_{q,c} \le v_{q,c} \le 1$. Therefore theorem 3 implies the abundance of good implementations with validity level sufficient for the error-free procedure. For the error-bounded procedure, O must satisfy $\left\|\Omega q_{1/2}\right\|_2^2 < 2\nu_q$. Such behavior may be achieved by designing O to yield value 1 at completion of each machine cycle, whereas leaving it uncontrolled ("floating") in-between, such that O has a sufficiently strong classical correlation with the states $q_{kT}$.

<u>Corollary 3</u>
In addition to the premises of theorem 3, let M be programmed such that $p > p_0$. Then any of the following implies that P is good with probability $1 - O(p_0^{-1})\,(p_0 \to \infty)$:

1. M is validable.
2. $\varepsilon > \frac{1}{2}$.

### 3.2.3   Time Dependent Hamiltonians

So far we have assumed that H is time-independent. On the other hand, current experimental approaches to the realization of quantum circuits maintain a coherent state and trigger state transitions with classical interventions (such as laser beams). The first quantum computers thus are likely to have time-dependent Hamiltonians.

Fortunately, instant computing works well when the Hamiltonian is turned on and off, assuming the form $f'(t)H\ (0 \le f'(t) \le 1)$ and yielding $U_t = e^{-i\frac{H}{\hbar}f\,t}$. The evolution can be stopped when $f(t) = \frac{1}{2}$, and the result can easily be measured.

## 3.3   Schrödinger Automata

<u>Definition 3.1</u>
A Schrödinger Automaton (SA) is a HA with Hamiltonian of Schrödinger form

$$H = -\Delta + V$$

Though the axioms of QM only demand a Hamiltonian to be self-adjoint, SA are "more physical" than mere HA, as many quantum-mechanical models require Schrödinger Hamiltonians. For several reasons, it is far more difficult to find implementations of QTMs into SA than into general type HA:

- The Hamiltonian cannot be constructed by adjoining eigenvalues and eigenfunctions one by one, because by any single eigenvalue and eigenfunction the potential and thus the whole Schrödinger operator is completely defined.
- Choice of the basis B determines orbits completely. But there are tuples of states that cannot be fitted by an orbit of any Schrödinger Hamiltonian (see proposition 3.1 below).
- The potential $V = x^2$ has the suitable spectrum, but eigenfunctions cannot be assigned arbitrarily. The only degree of freedom is in the mapping $\varphi$, but this must be simple and independent of the computational task.
- Aperiodic orbits are subject to the RAGE theorem [RS80] and other dynamical peculiarities [LA96, KL99, KKL01, Si90] of Schrödinger Hamiltonians. This imposes constraints on implementations of partial (non-total) QTMs which are difficult to handle.

The freedom of choice of basis B is further restricted by the following result.

<u>Proposition 3.1</u>
For any $n \geq 2, d \geq 1$ there are $h_k \in L^2(\mathbb{R}^d)$ $(k = 1, \ldots, n)$, which do not lie on one and the same orbit for any Schrödinger Hamiltonian.

Proof: Let for some real coefficients $\sum a_k |h_k|^2 = 0$, $\sum a_k = 0$, but $\sum a_k(h_k, -\Delta h_k) \neq 0$. Then for any $H = -\Delta + V$ : $\sum a_k(h_k, -\Delta h_k) \neq 0$, as $\sum a_k(h_k, Vh_k) = \sum a_k \int V|h_k|^2 = \int V \sum a_k |h_k|^2 = 0$. Would the functions lie on the same orbit, then they would all have the same energy $E$. But then $\sum a_k E = E \sum a_k = 0$, in contradiction to the previous equation.          q.e.d.

The same effect holds for $n \geq 3$ when including electromagnetic interaction.

By proposition 3.1, when B is chosen, then, for some computation, it may be impossible to find a Schrödinger Hamiltonian with an orbit fitting the image (under $\varphi$) of that computation. An important open question is thus the existence of QTM implementations into SA with simple $\varphi$, furthermore the existence of good SA implementations of totally periodic QTMs.

## 4    Conclusion

### 4.1    Summary and Open Problems

In the framework of non-relativistic, operational Quantum Mechanics, we have defined Hamiltonian Automata and physical implementations of Quantum Turing Machines. These definitions are natural and the HA a valid computational model. The definition of physical complexity (definition 1.4) has set the basis for a scale-independent gauge relating computational complexity of QTM and HA, and for the definition of efficient physical implementations.

Thereafter, we have defined error-free and error-bounded instant computation procedures, using repeated preparation and measurement. These are standard techniques in experimental physics, well-defined in QM. And it is also the way QTMs, if they could be realized, would be operated.

Finally, we have defined specific implementations of DTMs suitable for instant computation, proven that the results are obtained as expected, and explicitly constructed good implementations yielding the results with arbitrarily low error probabilities.

In theorem 1 we have then established the main results on the computational power of instant computation, shown in theorem 2 the mathematical existence of implementations with bounded physical resources and proven in theorem 3, that most implementations of QTMs are good in the sense of definition 2.2.

This all is standard QM and QTM theory. Our result relies only on the fact, that physical systems evolve at intermediate times, i.e. between two computational steps, into superpositions of the application of arbitrary powers of the Hamiltonian, by virtue of the Taylor expansion

$$e^{-iHt} = \sum \frac{(-iHt)^n}{n!}.$$

The eigenstates are superpositions of all states occurring in the course of a computation, thus anticipating computational results. Instant computation exploits these effects by implementing a reversible DTM in a QTM, preparing the initial state, starting the machine and measuring the state after half a machine cycle. As we have seen, this mode of operation opens a side channel.

Implementing a QTM into a physical system P requires to control the state of P at integer multiples of $T$. Should one day this technological challenge be mastered, then, as theorem 3 shows, these QTMs are likely to exhibit the instant computing effect when measuring the state at time $\frac{T}{2}$.

Instant computation does not impose any further constraints on implementations. All our results hold for arbitrary observable set O, or equivalently, mapping $\varphi$ or basis B. No account on real numbers is made, and no particularly high precision measurements nor any other capabilities beyond QTM technology are required.

Instant computation by a HA P may even be easier to realize than quantum operation by (the same or another) implementation, as in the former case the coherent quantum state of P needs to be maintained only for time $t_n < T$, whereas decoherence is a severe problem for long quantum computations. Instant computation may help to overcome the coherence problem.

We believe that instant computing is likely to become reality, under the proviso that QTMs can be built, which is the major open problem. Whereas theorem 3 is encouraging, section 3.3. raises issues regarding SA implementations needing further investigation. However, if QTMs can be realized, they will be SA. As well, the impacts of Quantum Field Theory (QFT) owe consideration.

Our results demonstrate differences between computation with physical devices and Turing computation. This opens a wealth of questions in mathematical logics, theory of computation, computational complexity and (complexity-based) cryptology.

## 4.2    Philosophical Digression

Since the days of Church and Turing, the Turing-machine (TM) is commonly considered the most general model of computation. The so-called Church-Turing thesis (CT) asserts according to [SEP], that any effective mechanical method that a *human* can carry out *unaided* by any machinery, save paper and pencil, can be computed by a TM.

In the course of time, generalizations of the CT far beyond its original form have become popular among scientists. Dropping the word "unaided", any problem solvable by some physical computing device is now claimed to be Turing-computable. It is postulated that any machine can be simulated by a universal TM. Finally the brain, and indeed any biological or physical system is assumed to be susceptible to TM simulation. The so-called Extended CT (ECT) even assumes that such simulation, when using a device of suitable architecture and technology, can be effected with at most a polynomial slow-down.

Recently, the generalized form CT, and ECT have been challenged by some authors such as [Akl05, CP06, Si95, Yao06], challenging the realizability of universal Turing Machines [Akl05] or arguing that physical systems might be capable to solve Turing-uncomputable problems, or to solve Turing-computable problems faster than any Turing Machine.

Our findings show that the latter is indeed the case:

Proposition 4.1
a) There exist physical processes of arbitrary computational time complexity.
b) There exist Turing-uncomputable physical processes.
c) The ECT is wrong.
d) The generalizations of the CT mentioned above are wrong.
e) The recognition of any recursively enumerable set, and its complement, is in HBPL.
f) The halting problem is in HBPL.

Proof:

a) The state of the HA of theorem 1a) above at time $\frac{1}{2}$ is at least as hard to compute by a Turing Machine as the underlying problem.

b) The state of the HA of theorem 1b) above at time $\frac{1}{2}$ is Turing-uncomputable.

c) This is a direct consequence of theorem 1a), as well as of theorem 1b).
d) This is a direct consequence of theorem 1b).
e) This is a direct consequence of theorem 1b).
f) This is a direct consequence of theorem 1b).

The generalizations of the CT, and the ECT have given rise to many "in principle" arguments and led to assertions such as that the brain were a Turing Machine "in principle", or natural language did possess a finite generating grammar "in principle", and any biological and physical system could be efficiently digitally simulated "in principle". The ECT has awarded the status of a law of conservation of computational time complexity, almost as fundamental as the energy conservation law.

This has narrowed the conceptual space in the reasoning about many natural phenomena and barred routes potentially leading to new insights. As these bonds are widened, one may expect progress in artificial intelligence and other disciplines. Many biological, mental and societal processes can be characterized as kind of *collective information processing*, by cells, neurons, animals and humans.

The capability to solve the halting problem refutes the mentioned generalizations of CT, and a fortiori the ECT. Instant computing and related physical effects may be exploited in future computing devices as well as in tools and engines designed for physical manipulation, and it may well be the case that nature makes use of this and other effects. The fall of the CT may allow a new understanding of collective information processing in biological, mental and social systems. The power of physical processing will be a promising, rich field in future research.

Concluding, we believe that our result provides some evidence for something different from, incommensurable to and superior over Turing computation. QM allows for effects beyond those exploited by QTMs, enabling physical devices to outperform TMs and QTMs.

Other, even more powerful effects and variations of *physical processing* may still await discovery. New insights in natural, societal and mental processes may result when studying them free from the generalized CT and ECT dogma. Yet another promising research field may be *physical manipulation*, i.e. the way man manipulates and forms matter and nature, using tools and physical, chemical and biological processes.